\documentclass[preprint,showkeys,showpacs,preprintnumbers,amsmath,amssymb,aps]{revtex4}
\usepackage{amsmath}
\usepackage{txfonts}

\usepackage{graphicx}
\usepackage{rotating}
\usepackage{dcolumn}
\usepackage{bm}

\begin{document}


\title{From nothing to something II:\\ nonlinear systems via consistent correlated bang}
\author{S Y Lou$^{1,2}$}
\affiliation{$^{1}$\footnotesize \it Ningbo Collabrative Innovation Center of Nonlinear Harzard System of Ocean and Atmosphere\\
\footnotesize \it and Faculty of Science, Ningbo University,  Ningbo, 315211, China,\\
\footnotesize $^{2}$
\it Shanghai Key Laboratory of Trustworthy Computing, East China Normal University, Shanghai 200062, China}

\date{\today}

\begin{abstract}
Chinese ancient sage Laozi said everything comes from \emph{\bf \em ``nothing"}. \rm In the first letter (Chin. Phys. Lett. 30 (2013) 080202), infinitely many discrete integrable systems have been obtained from ``nothing" via simple principles (Dao). In this second letter,  a new idea, the consistent correlated bang, is introduced to obtain nonlinear dynamic systems including some integrable ones such as the continuous nonlinear Schr\"odinger equation (NLS), the (potential) Korteweg de Vries (KdV) equation, the (potential) Kadomtsev-Petviashvili (KP) equation and the sine-Gordon (sG) equation. These nonlinear systems are derived from nothing via suitable ``Dao", the shifted parity, the charge conjugate, the delayed time reversal, the shifted exchange, the shifted-parity-rotation and so on.
\end{abstract}



\maketitle

\bf \em 1. Introduction. \rm Around the 6th century BC, in chapter 42 of a Chinese classical text `\emph{Daodejing}', the author sage Laozi said `\emph{Dao sheng yi, yi sheng er, er sheng san, san sheng wanwu, $\cdots$}' \cite{TaoTeJing}. The correct translation of the Laozi's ideology should have the form \cite{LLT}, ``Nothing produces the first via `Dao', then the first produces the second, the second produces the third, and the third produces  everything, ...". In other words, the essence of the Laozi's philosophy is that everything comes from nothing through a suitable ``Dao"! Clearly, the Big Bang Theory on our universe and the Dirac sea in quantum physics are two of the best manifestations of the Laozi's philosophy. 
In Ref. \cite{LLT} Laozi's idea has been really used to produce infinitely many discrete integrable models from ``nothing". 

A natural important question has been left in the first paper: \\ \em 
How to find continuous models including some integrable models directly from ``nothing" via suitable ``Dao"? \\
\rm
To partially answer this question, we will introduce a new idea called ``consistent correlated bang" (CCB) which splits one trivial  equation (``nothing") to two or more equations while the split equations are correlated and consistent. It will be interesting that the nontrivial equation systems such as the well known nonlinear Schr\"odinger equation (NLS), (potential) Korteweg de Vries (KdV) equation, (potential) Kadomtsev-Petviashvili (KP) equation,  sine-Gordon (sG) equation and so on may be naturally included in the consistent and correlated banged systems. 

\bf \em 2. General CCB. \rm As in \cite{LLT},  mathematically, the ``nothing" can be written as $0=0$, or for convenience later, we write it as
\begin{equation}
L \psi=0,\ \psi=0, \label{Lpsi}
\end{equation}
where $L$ may be simply taken as a linear operator. Different choices of $L$ in \eqref{Lpsi} will related to different selections of ``Dao". 

To obtain some nontrivial systems from trivial relation \eqref{Lpsi}, we take three steps, (i) Bang, (ii) Correlation and (iii) Consistence. 

(i) \em Bang. \rm Firstly, we 
write $\psi =\sum_{i=1}^n \psi_i$ and the right $0$ of \eqref{Lpsi} as $0=\sum_{i=1}^n G_i$. Thus, \eqref{Lpsi} becomes 
\begin{equation}
L\sum_{i=1}^n \psi_i=\sum_{i=1}^n G_i,\ \ \sum_{i=1}^n \psi_i=0, \ \ \sum_{i=1}^n G_i=0.\label{Lpsi1}
\end{equation}
It is clear that the trivial equation \eqref{Lpsi} can be banged to an equation set 
\begin{equation}
\Delta\equiv\{\Delta_i(\equiv L \psi_i-G_i=0),\ \ i=1,\ 2,\ \ldots,\ n\}
\label{Lpsi2}
\end{equation}
with
\begin{equation}
 \sum_{i=1}^n \psi_i=0, \ .\label{Lpsi01}
\end{equation}
and 
\begin{equation}
 \sum_{i=1}^n G_i=0.\label{Lpsi02}
\end{equation} 

(ii) \em Correlation. \rm 
Because $\psi_i$ and $G_i$ are banged from $\psi(=0)$ and $0$ respectively, it is reasonable  that there are some relations among them. In other words they are correlated each other and may be connected by 
\begin{equation}
 \psi_i=\hat{f}_i\psi_1,\quad 
  G_i=\hat{g}_iG_1 \label{Lpsi00}
\end{equation} 
for suitable operators $\{f_i,\ g_i\},\ i=1,\ 2,\ \ldots,\ n$. 
  
(iii) \em Consistency. \rm 
Because of the correlated condition \eqref{Lpsi00}, we can find some necessary consistent conditions. 

The first set of consistent conditions possesses the form 
\begin{equation}
g_iL=Lf_i, \ i=1,\ 2,\ \ldots,\ n
  \label{fg}
\end{equation} 
which can be directly obtained by applying $\hat{g}_i$ on 
\begin{equation}
 \Delta_1\equiv L\psi_1-G_1=0.
  \label{Eq1}
\end{equation} 

The second type of consistent conditions read 
\begin{equation}
\hat{f}_i\Delta =\Delta,\ i=1,\ 2,\ \ldots,\ n,  \label{dFg}
\end{equation}
and 
\begin{equation}
\hat{g}_i\Delta =\Delta,\ i=1,\ 2,\ \ldots.\ n  \label{dgF}
\end{equation}
The conditions \eqref{dFg} and \eqref{dgF} can be equivalently expressed that 
 the sets
\begin{equation}
{\cal{G}}=\{\hat{g}_i,\ i=1,\ 2,\ \ldots,\ n\}, \label{calg}
\end{equation}
and
\begin{equation}
{\cal{F}}=\{\hat{f}_i,\ i=1,\ 2,\ \ldots,\ n \ \} \label{F}
\end{equation}
are all groups with $n$ elements. 

Substituting \eqref{Lpsi00} into \eqref{Lpsi02} leads to the third type of consistent condition 
\begin{equation}
\sum_{i=1}^ng_iG_1=0. \label{gG}
\end{equation}

Under the above consistent conditions, only one equations of $\Delta$, say, \eqref{Eq1} is independent while others can be reproduced by applying $\hat{f}_i$ or $\hat{g}_i$ on it. 

\bf \em 3. Nonlinear systems from CCB. \rm To show some concrete examples, we restrict the special cases for $n=2$. 
In this case, the only second order group is the cyclic group that means 
\begin{equation}
{\cal{F}}=\{\hat{f},\ I(=\hat{f}^2) \},\ \quad {\cal{G}}=\{\hat{g},\ I(=\hat{g}^2) \},
 \label{gFgG}
\end{equation}
where $I$ is the identity operator. 

The banged system in the $n=2$ situation becomes 
\begin{eqnarray}
\Delta_1&\equiv &L\psi_1-G_1=0,\label{d1}\\
\Delta_2&\equiv &L\psi_2-G_2=0,\label{d2}
\end{eqnarray}
while the correlated conditions are 
\begin{eqnarray}
&&\psi_2=\hat{f}\psi_1,\\
&& G_2=\hat{g}G_1.\label{cc}
\end{eqnarray}
The consistent conditions are simplified to 
\begin{eqnarray}
&&\hat{g}L =L\hat{f},
\label{gLLf}\\
&& \hat{f}^2=\hat{g}^2=1.\label{f2g2}
\end{eqnarray}
and 
\begin{equation}
G+gG=0,\ (G_1\rightarrow G). \label{cc1}
\end{equation}
For simplicity later, we take 
\begin{eqnarray}
\hat{g}=\delta \hat{f},\ \quad \delta^2=1. \label{rg}
\end{eqnarray}
Substituting 
\eqref{rg} into the consistent condition \eqref{gLLf}, we have 
\begin{eqnarray}
&&\hat{f}L\hat{f}^{-1} =\delta L
\label{fLLf}
\end{eqnarray}
that means $ \hat{f}$ is a symmetry transformation of $L$ for $\delta =1$ and antisymmetry transformation for $\delta=-1$. 

Now, we can select some possible $L$ to derive non-trivial models. 

\em A. NLS and derivative NLS equations. \rm To derivative the NLS equation, we select $L$ possessing the form
\begin{equation}
L=i\partial_t+\partial_x^2+\partial_y^2+\partial_z^2=i\partial_t+\nabla^2, \label{NLSL}
\end{equation}  that means \eqref{Lpsi} is an trivial linear Schr\"dinger equation without potential 
\begin{eqnarray}
i\psi_t+\nabla^2\psi=0. \label{LS}
\end{eqnarray}
After writing $\psi=\psi_1+\psi_2$ and $0=G-G$ ($G_1=G,\ G_2=-G$), the linear Schr\"odinger equation \eqref{LS} can be banged to 
\begin{eqnarray}
i\psi_{1t}+\nabla^2\psi_{1}=G. \label{LS1}
\end{eqnarray}
and 
\begin{eqnarray}
i\psi_{2t}+\nabla^2\psi_{2}=-G. \label{LS2}
\end{eqnarray}
For the given operator \eqref{NLSL}, its possible symmetry transformations with $\hat{f}^2=1$ can be selected as 
\begin{equation}
\hat{f}=\hat{T}_i\hat{P}_j,\quad i=1,\ 2,\ j=1,\ 2,\ 3,\ 4, \label{fij1}
\end{equation}
with 
\begin{equation}
\hat{T}_1t=t,\ \hat{T}_2= \hat{C}\hat{T}_d,\ \hat{C}\phi=\phi^*,\ \hat{T}_dt=-t+t_0,\label{Ti1}
\end{equation}
\begin{equation}
\hat{P}_1 \left(\begin{array}{l}
x\\ y\\ z
\end{array}\right)=\left(\begin{array}{lll}
1 & 0 & 0 \\ 0& 1 & 0\\ 0& 0 & 1
\end{array}\right)\left(\begin{array}{l}
x\\ y\\ z
\end{array}\right),\label{Pj1}
\end{equation}
\begin{equation}
\hat{P}_2 \left(\begin{array}{l}
x\\ y\\ z
\end{array}\right)=\left(\begin{array}{lll}
-1 & 0 & 0 \\ 0& -1 & 0\\ 0& 0 & -1
\end{array}\right)\left(\begin{array}{l}
x\\ y\\ z
\end{array}\right)+\left(\begin{array}{l}
x_0\\ y_0\\ z_0
\end{array}\right),\label{Pj2}
\end{equation}
\begin{equation}
\hat{P}_3 \left(\begin{array}{l}
x\\ y\\ z
\end{array}\right)=\left(\begin{array}{ccc}
1-(1+\sin\theta) \cos^2\gamma & -\sin\gamma\cos\gamma(1+\sin\theta) & \cos\theta\cos\gamma
 \\ -\sin\gamma\cos\gamma(1+\sin\theta) & 
 (1+\sin\theta) \cos^2\gamma-\sin\theta & \cos\theta\sin\gamma\\ \cos\theta\cos\gamma& \cos\theta\sin\gamma & \sin\theta
\end{array}\right)\left(\begin{array}{c}
x\\ y\\ z
\end{array}\right)+\left(\begin{array}{c}
z_0\cos\theta\cos\gamma\\ z_0\cos\theta\sin\gamma\\ -z_0
\end{array}\right),\label{Pj3}
\end{equation}
\begin{equation}
\hat{P}_4 \left(\begin{array}{l}
x\\ y\\ z
\end{array}\right)=\left(\begin{array}{ccc}
(1-\sin\theta) \cos^2\gamma-1 & \sin\gamma\cos\gamma(1-\sin\theta) & \cos\theta\cos\gamma
 \\ \sin\gamma\cos\gamma(1-\sin\theta) & 
 (\sin\theta-1) \cos^2\gamma-\sin\theta & \cos\theta\sin\gamma\\ \cos\theta\cos\gamma& \cos\theta\sin\gamma & \sin\theta
\end{array}\right)\left(\begin{array}{c}
x\\ y\\ z
\end{array}\right)+\left(\begin{array}{c}
x_1\\ -y_0\\ -z_0
\end{array}\right),\label{Pj4}
\end{equation}
where $x_1=y_0\sin\gamma+z_0(1+\sin\theta)$,\ $\theta,\ \gamma,\ y_0,\ t_0$ and $z_0$ are arbitrary constants. 

$\hat{T}_1$ and $\hat{P}_1$ are identity transformations for the time and space respectively. $\hat{C}$ is a charge conjugate operator. $\hat{T}_d$ is a delayed time reversal operator. $\hat{P}_2$ is a shifted parity operator for all space component. 
$\hat{P}_3$ is a combination of the rotation and the shifted parity for one space component while $\hat{P}_4$ is a combination of the rotation and the shifted parity for two space component. 

Applying the $\hat{f}$ operator \eqref{fij1} with the condition $\hat{f}\psi_2=\psi_1$ on \eqref{LS2} and comparing the result with \eqref{LS1}, we have the consistent condition for $G$,
\begin{equation}
\hat{f}G=-G \label{cG1}
\end{equation} 
which is same as \eqref{cc1}.

For any given $\psi_1$-independent $G=G(x,\ y,\ z,\ t)$ with \eqref{cG1}, the banged system is trivially integrable due to its linearity. 

It is interesting that the consistent condition \eqref{cG1} permits various nonlinear realizations. 
For instance, if we take 
\begin{equation}
\hat{f}=\hat{C} \hat{T}_d
\hat{P}_2, 
\end{equation}
we can take  ($x_1=x,\ x_2=y,\ x_3=z$)
\begin{equation}
G=\alpha (\psi_1^2\psi_2^*-\psi_2^2\psi_1^*)+i\sum_{j=1}^3\beta_j \left\{\psi_1\psi_{1x_j} \psi_2^*-\psi_2\psi_{2x_j} \psi_1^*\right\}.\label{G1}
\end{equation}
Thus, the banged system \eqref{LS1} becomes 
\begin{equation}
i\psi_{1t} +\nabla^2 \psi_1=\alpha (\psi_1^2\psi_2^*-\psi_2^2\psi_1^*)+i\sum_{j=1}^3\beta_j \left\{\psi_1\psi_{1x_j} \psi_2^*-\psi_2\psi_{2x_j} \psi_1^*\right\}.\label{CNLS1}
\end{equation}
Finally, substituting $\psi_2=-\psi_1$ into \eqref{CNLS1}, we have 
\begin{equation}
i\psi_{1t} +\nabla^2 \psi_1+\alpha |\psi_1|^2\psi_1+i\sum_{j=1}^3\beta_j |\psi_1|^2\psi_{1x_j}=0.\label{GNLS1}
\end{equation}
Equation \eqref{GNLS1} is a generalization of the nonlinear Schr\"dinger (NLS) equation ($\beta_j=0$) and the derivative NLS (DNLS) equation. The NLS and the DNLS systems are integrable only in 1+1 dimensional case. 

\em B. KdV, potential KdV and modified KdV equations. \rm

If the operator $L$ is fixed as
\begin{equation}
L=\partial_t+\partial_x^3,\label{KdVL}
\end{equation}
 then one can find possible antisymmetry ($\delta=-1$ in \eqref{fLLf}) operators 
of $L$ in the forms 
\begin{equation}
\hat{f}=\hat{C}^a\hat{T}_d\hat{P}^x_s, \quad a=0\ \mbox{\rm or}\ 1 \label{kdvf}
\end{equation}
where $\hat{C}, \ \hat{T}_d$ and $\hat{P}^x_s$ are charge conjugate, delayed time reversal and $x$-space shifted parity operators respectively. 

For the antisymmetry operator(s) \eqref{kdvf}, the consistent condition \eqref{cc1} becomes 
\begin{equation}
\hat{f}G=G, \label{kdvG}
\end{equation}
and the banged system has the form
\begin{equation}
\psi_{1t}+\psi_{1xxx}=G.  \label{kdvpsi1}
\end{equation}

Similar to the NLS case, there are various possible nonlinear realizations of \eqref{kdvG}. Here is a possible simple differential polynomial realization 
\begin{equation}
G=\alpha (\psi_1^2\psi_{2x}-\psi_2^2\psi_{1x})-\beta \psi_{1x}\psi_{2x}. \label{kdvG1}
\end{equation}
Substituting \eqref{kdvG1} into \eqref{kdvpsi1}, we have 
\begin{equation}
\psi_{1t}+\psi_{1xxx}+\alpha(\psi_2^2\psi_{1x}-\psi_1^2\psi_{2x}) +\beta \psi_{1x}\psi_{2x} =0. \label{kdve}
\end{equation}

By using the trivial solution $\psi_2=-\psi_1=-u$ yields 
\begin{equation}
u_{t}+u_{xxx}+2\alpha u^2u_{x}-\beta u_{x}^2=0 \label{kdve1}
\end{equation}
which is a generalization of the potential KdV ($\alpha=0$) and modified KdV ($\beta=0$). The KdV equation can not be directly obtained from the realization of \eqref{kdvG} though it can be considered as simple transformations from the potential KdV and the modified KdV systems. 

\em C. Nonlinear Klein-Gordon systems. \rm 

To find nonlinear 1+1 dimensional Klein-Gordon equation systems, we can use the following $L$ operator
\begin{equation}
L=\partial_t^2-\partial_x^2. \label{KGL}
\end{equation}
Higher dimensional nonlinear Klein-Gordon equation systems can also be obtained by using the transformation $\partial_x^2 \rightarrow \nabla^2$. 

For the operator \eqref{KGL}, its symmetry operator $\hat{f}$ with $\hat{f}^2=1$ possesses the form
\begin{equation}
\hat{f}_s=\hat{C}^{a_1}\hat{T}_d^{a_2} \hat{P}_s^{a_3}\hat{R}^{a_4}\left(\hat{E}_1\hat{E}_2 \right)^{a_5},\ a_i=0 \ \mbox{\rm or}\ 1,\ i=1,\ 2,\ \ldots 5, \label{KGfs}
\end{equation}
where $\hat{C}$ is the charge conjugate, $\hat{T}_d$ is the delayed time reversal ($\hat{T}_d t=-t+t_0$), 
$\hat{P}_s$ is the shifted $x$-parity ($\hat{P}_s x=-x+x_0$), $\hat{R}$ is the shifted-parity-rotation operator defined by ($\delta^2=\delta_1^2=1$)
\begin{equation}
\hat{R}\left(\begin{array}{c}
x\\ t
\end{array}\right)
=\left(\begin{array}{cc}
-\delta \cosh \theta & \delta_1 \sinh \theta\\ -\delta_1 \sinh \theta & \delta\cosh \theta
\end{array}\right)\left(\begin{array}{c}
x\\ t
\end{array}\right)
+\left(\begin{array}{c}
x_1(1+\delta\cosh \theta)\\ \delta_1\sinh \theta x_1
\end{array}\right),
\end{equation}
$\hat{E}_1$ is the shifted exchange with the definition 
\begin{equation}
\hat{E}_1\left(\begin{array}{c}
x\\ t
\end{array}\right)=\left(\begin{array}{cc}
0 & 1\\ 1 & 0
\end{array}\right)\left(\begin{array}{c}
x\\ t
\end{array}\right)+\left(\begin{array}{c}
x_2 \\ -x_2
\end{array}\right)
\end{equation}
and $\hat{E}_2$ is the shifted exchange parity with 
\begin{equation}
\hat{E}_2\left(\begin{array}{c}
x\\ t
\end{array}\right)=\left(\begin{array}{cc}
0 & -1\\ -1 & 0
\end{array}\right)\left(\begin{array}{c}
x\\ t
\end{array}\right)+\left(\begin{array}{c}
x_3 \\ x_3
\end{array}\right)
\end{equation}
while $t_0,\ x_i,\ i=0,\ 1,\ 2,\ 3$ and $\theta$ are arbitrary constants. 

For the operator \eqref{KGL}, there are not only the symmetry operators \eqref{KGfs}, but also the antisymmetry operators with the form ($\hat{f}\rightarrow \hat{f}_a$)
\begin{equation}
\hat{f}_a=\hat{C}^{a_1}\hat{T}_d^{a_2} \hat{P}_s^{a_3}\hat{R}^{a_4}\hat{E}_j,\ a_i=0 \ \mbox{\rm or}\ 1,\ i=1,\ 2,\ \ldots 4,\quad j=1 \ \mbox{\rm or}\ 2. \label{KGfa}
\end{equation}

For the operator \eqref{KGL}, the banged system possesses the form 
\begin{equation}
\psi_{1tt}-\psi_{1xx}=G
\end{equation}
with the consistent conditions
\begin{equation}
\hat{f}_sG=-G,\label{KGs}
\end{equation}
or 
\begin{equation}
\hat{f}_aG=G. \label{KGa}
\end{equation}
For the consistent conditions \eqref{KGs} and \eqref{KGa} we select the nonlinear realizations 
\begin{equation}
G=\alpha\sin (\psi_1-\psi_2)+\beta_1 (\psi_1-\psi_2)+\beta_2 (\psi_1-\psi_2)^3, \label{KGGs}
\end{equation}
and 
\begin{equation}
G=K(\psi_1-\psi_2),\quad K(-u)=K(u),\label{KGGa}
\end{equation}
respectively, where $K$ is an even function of a single argument $\psi_1-\psi_2$. Under the selections \eqref{KGGs} and \eqref{KGGa}, the related banged systems possesses the forms 
\begin{equation}
\psi_{1tt}-\psi_{1xx} =\alpha\sin (\psi_1-\psi_2)+\beta_1 (\psi_1-\psi_2)+\beta_2 (\psi_1-\psi_2)^3, \label{sGphi4}
\end{equation}
and 
\begin{equation}
\psi_{1tt}-\psi_{1xx} =K(\psi_1-\psi_2), \label{KcG}
\end{equation}
respectively. 
With help of the trivial solution condition $\psi_2=-\psi_1=-u$, we have 
\begin{equation}
u_{tt}-u_{xx} =\alpha\sin (2u)+2\beta_1 u+8\beta_2 u^3, \label{sGu}
\end{equation}
and 
\begin{equation}
u_{tt}-u_{xx} =K(2u). \label{KcGu}
\end{equation}
It is clear that \eqref{sGu} is a generalization of the well known sine-Gordon ($\beta_1=\beta_2=0$) and $\phi^4$ ($\alpha=0$) models. If $K(2u)$ is taken as $K(2u)\sim \cos(ku)$, \eqref{KcGu} is also equivalent to a sine-Gordon equation after a simple field shift ($ku\rightarrow ku+\pi/2$). 

\em D. Potential KP equation. \rm 
In order to derivative the potential KP equation, we can take a linear operator with weak two dimensional approximation 
\begin{equation}
L=\partial_t\partial_x+\partial_x^4+\partial_y^2. \label{KPL}
\end{equation}
It is clear that the operator $L$ given in \eqref{KPL} possesses the symmetry operator 
\begin{equation}
\hat{f}=\hat{C}^{a_1}\hat{P}_{ys}^{a_2}(\hat{T}_d\hat{P}_{xs})^{a_3},\quad a_i=0 \ \mbox{\rm or}\ 1,\quad
i=1,\ 2,\ 3.
\end{equation}
In this case the consistent banged system possesses the form
\begin{equation}
\psi_{1xt}+\psi_{1xxxx}+\psi_{1yy}=G,\quad \hat{G}=-G.
\end{equation}
The following nonlinear realization 
\begin{equation}
G=-\alpha(\psi_{1x}\psi_{2x})_x
\end{equation}
leads to the potential KP equation ($\psi_{2}=-\psi_1=-u$)
\begin{equation}
(u_{t}+u_{xxx}+\alpha u_x^2)_x+u_{yy}=0. \label{pKP}
\end{equation}

\bf \em 4. Summary and discussions. \rm 

In summary, a large number of nonlinear systems, including some possible integrable ones such as the well known NLS, (potential) KdV, (potential) KP and sine-Gordon systems may be derived from nothing (trivial $0=0$ equation) by introducing the new idea ``Consistent Correlated Bang" and other suitable ``Dao" which are implied by some symmetric or antisymmetric operators. The main operators we introduced are the time-dependent Schr\"odinger operator \eqref{NLSL} without potential, the linear dispersive wave operator \eqref{KdVL} without dissipative, the linear wave operator (the operator of boson field without mass) \eqref{KGL}, the dispersive wave operator with weak two dimensional approximation \eqref{KPL} and so on. These operators possesses some types of symmetries. The main symmetries are charge conjugate, (delayed-) time reversal, (shifted-) parity, (shifted-parity-) rotations, (shifted-) exchanges and so on. 

In this letter, some single component nonlinear systems are derived. In fact, multiple component nonlinear systems may also be obtained by means of the similar way with help of the finite dimensional symmetry group which are related to some suitable multicomponent operators.

Furthermore, the idea used here can also be applied for nontrivial systems to get more interesting nonlinear models which are defined as Alice-Bob systems \cite{AB}. The Alice-Bob systems may be used to describe multi-place physics \cite{AB1}. 

\section*{Acknowledgement}
The author are in debt to the helpful discussions with Professors Hu X. B., Liu Q. P.,  D. J. Zhang and Y. Chen.
The work was sponsored by the Global Change Research
Program of China (No.2015CB953904), the National Natural Science Foundations of China (Nos. 11435005), Shanghai Knowledge Service Platform for Trustworthy Internet of Things (No. ZF1213) and K. C. Wong Magna Fund in Ningbo University.


\begin{thebibliography}{99}
\bibitem{TaoTeJing}Boltz W G 1993 Lao tzu Tao te ching. In Early Chinese Texts: A Bibliographical Guide, edited by Michael Loewe. Berkeley: University of California, Institute of East Asian Studies pp. 269-92
\bibitem{LLT} Lou S Y, Li Y Q and Tang X Y 2013 Chin. Phys. Lett. 080202
\bibitem{appl}  Crighton D G 1995,
Acta Appl. Math. 39 39
\bibitem{AB}Lou S Y 2016 \em Alice-Bob systems, $P_s-T_d-C$ principles and multi-soliton solutions, \rm arXiv: 1603.03975v2. nlin.SI. 
\bibitem{AB1}Lou S Y  and Huang F 2016 \em  Alice-Bob Physics: Coherent Solutions of Nonlocal KdV Systems, \rm arXiv: 1606.0154v2. nlin.SI. 
\end{thebibliography}
\end{document}